\documentclass[sigconf,natbib=false]{acmart}

\AtBeginDocument{%
  }

\setcopyright{acmcopyright}
\copyrightyear{2023}
\acmYear{2023}
\acmDOI{XXXXXXX.XXXXXXX}
\acmConference[SC23]{SuperComputing 2023}{November 2023}{Denver, Colorado USA}
\copyrightyear{2023}

\RequirePackage[
  datamodel=acmdatamodel,
  style=acmnumeric,
  sorting=none
  ]{biblatex}

\usepackage{caption}
\usepackage{subcaption}

\usepackage{lscape}
\usepackage{adjustbox}

\usepackage{enumitem}
\setlist{leftmargin=4mm}

\usepackage{glossaries}
\glsdisablehyper
\newacronym{SIF}{SIF}{Singularity Image Format}
\newacronym{WLM}{WLM}{Workload Manager}
\newacronym{CNCF}{CNCF}{Cloud Native Computing Foundation}
\newacronym{GCP}{GCP}{Google Cloud Platform}
\newacronym{SBOM}{SBOM}{Software Bill of Materials}
\newacronym{VMM}{VMM}{Virtual Machine Monitor}
\newacronym{UserNS}{\texttt{UserNS}}{user namespace}
\newacronym{OCI}{OCI}{\emph{Open Container Initiative}}
\newacronym{HPC}{HPC}{High-performance computing}

\makeatletter
\newcommand{\pkg}{\@ifstar{\pkg@star}{\pkg@nostar}}
\newcommand{\pkg@star}[1]{\texttt{#1}}
\newcommand{\pkg@nostar}[1]{\texttt{#1}~\citex{#1}}
\newcommand{\citex}[1]{%
    \expandafter\ifx\csname cit:#1\endcsname\relax
        \expandafter\gdef\csname cit:#1\endcsname{}%
        \cite{#1}%
    \fi
}
\newcommand{\fpath}[1]{\texttt{#1}}
\makeatother

\begin{document}


\title{Survey of adaptive containerization architectures for HPC}
% The title of your work should use capital letters appropriately -
% \url{https://capitalizemytitle.com/} has useful rules for
% capitalization. Use the {\verb|title|} command to define the title of
% your work. If your work has a subtitle, define it with the
% {\verb|subtitle|} command.  Do not insert line breaks in your title.

% If your title is lengthy, you must define a short version to be used
% in the page headers, to prevent overlapping text. The \verb|title|
% command has a ``short title'' parameter:
% \begin{verbatim}
%   \title[short title]{full title}
% \end{verbatim}

\author{Tiziano Müller}
\authornote{Both authors contributed equally to this research.}
\email{tiziano.mueller@hpe.com}
\author{Nina Mujkanovic}
\authornotemark[1]
\email{nina.mujkanovic@hpe.com}
\affiliation{%
  \institution{HPE HPC/AI EMEA Research Lab}
  \streetaddress{Innere Margarethenstrasse 5}
  \city{Basel}
  \state{BS}
  \country{Switzerland}
  \postcode{4051}
}

\author{Juan J. Durillo}
\email{juan.durillobarrionuevo@lrz.de}
\author{Nicolay Hammer}
\email{nicolay.hammer@lrz.de}
\affiliation{%
  \institution{Leibniz Supercomputing Center}
  \streetaddress{Boltzmannstraße 1}
  \city{Munich}
  \state{Bavaria}
  \country{Germany}
}


\begin{abstract}
  Containers offer an array of advantages that benefit research reproducibility and portability across groups and systems.
  As container tools mature, container security improves, and \gls{HPC} and cloud system tools converge, supercomputing centers are increasingly integrating containers in their workflows.
  The technology selection process requires sufficient information on the diverse tools available, yet the majority of research into containers still focuses on cloud environments.
  
  We consider an adaptive containerization approach, with a focus on accelerating the deployment of applications and workflows on \gls{HPC} systems using containers.
  To this end, we discuss the specific \gls{HPC} requirements regarding container tools, and analyze the entire containerization stack, including container engines and registries, in-depth.
  Finally, we consider various orchestrator and \gls{HPC} workload manager integration scenarios. 
  
\end{abstract}


\begin{CCSXML}
<ccs2012>
   <concept>
       <concept_id>10010520.10010521.10010537.10003100</concept_id>
       <concept_desc>Computer systems organization~Cloud computing</concept_desc>
       <concept_significance>500</concept_significance>
       </concept>
   <concept>
       <concept_id>10011007.10010940.10010941.10010949</concept_id>
       <concept_desc>Software and its engineering~Operating systems</concept_desc>
       <concept_significance>300</concept_significance>
       </concept>
   <concept>
       <concept_id>10002978.10003006.10003007</concept_id>
       <concept_desc>Security and privacy~Operating systems security</concept_desc>
       <concept_significance>500</concept_significance>
       </concept>
 </ccs2012>
\end{CCSXML}

\ccsdesc[500]{Computer systems organization~Cloud computing}
\ccsdesc[300]{Software and its engineering~Operating systems}
\ccsdesc[500]{Security and privacy~Operating systems security}

\keywords{Containers,HPC,High performance computing,Kubernetes,Survey}

%\received{TODO}

\maketitle

\section{Introduction}

The building blocks for containers were laid during the development of Unix V7 in 1979, when chroot - the possibility to change the root directory of a process and its children - was introduced. The design of cgroups or "Control Groups" in 2006 added the possibility for limiting, accounting, and isolating resource usage of a collection of processes. In 2008, using chroot, cgroups, and Linux namespaces, LXC (LinuX Containers) were implemented as the first, most complete version of what we today consider a standard container manager. It wasn't until the development of a container engine - Docker, in 2013 - and cloud computing systems that containers truly exploded in popularity.

Since then, as container tools matured, container security improved, and \gls{HPC} and cloud system tools converged, supercomputing centers have integrated containers in their workflows \cite{benedicic2016opportunities, yin2019strategies, younge2021constructing}. Containers offer an array of advantages, such as greater efficiency than full hardware-level virtualization, support for DevOps, management of on-demand resources, scalability across resources, and improved portability of workflows. All of this in turn enhances the reproducibility of scientific research, as well as the cooperation between researchers.

Research on containers focused mostly on cloud solutions \cite{zhouContainerisationHighPerformance2023}, orchestration in cloud environments \cite{beltreEnablingHPCWorkloads2019}, and benchmarking containers running in cloud systems vs bare metal \cite{kellertesserContainersHPCSurvey2022}. With the emergence of more heterogeneous systems and convergence between \gls{HPC}, cloud, and edge, more research has been performed on the use and integration of container technologies and orchestrators in \gls{HPC} environments \cite{georgiou2020converging, younge2017twosystems}.

In this paper, we perform an in-depth analysis of container technologies with a focus on \gls{HPC}, integration into an \gls{HPC} environment, performance, and ease-of-use. Further, we describe integration scenarios taking the limitations of \gls{HPC} sites into account. Finally, we consider an orchestrator integration scenario, specifically that of merging Kubernetes into an \gls{HPC} workload management environment, with a proof of concept of a possible Kubernetes-in-Slurm integration.

We suggest an approach to portable but performant containers in \gls{HPC} that we term \emph{adaptive containerization}. Adaptive containerization focuses on accelerating the deployment of applications and workflows using containers while distributing some of the maintenance burden and providing additional possibilities to the research scientist. Essentially, it includes the integration of \gls{HPC}-centric and specific container engines, registries, and orchestration tools, to deliver full workflow capabilities to an end user.

To our knowledge, while there have been papers with the goal of creating taxonomies over all container applications available for the cloud or \gls{HPC}/cloud environment \cite{casalicchio2020state, pahl2019review}, empirical analyses of select aspects of container technologies \cite{abraham2020containers}, and surveys delving into the similarities and differences between containers deployed on cloud and \gls{HPC} environments \cite{zhouContainerisationHighPerformance2023}, this is the first analysis of the entire containerization, registry, and orchestration stack with a purely \gls{HPC}-centric focus. 

The paper is organized as follows: section~\ref{sec:motivation} discusses the motivation to this work, in section~\ref{sec:concepts} we introduce relevant terminology, as well as \gls{HPC} specific requirements to containerization. An in-depth analysis of container engines and container registries for \gls{HPC} is performed in sections \ref{sec:container-engine-comparison} and \ref{sec:registry-comparison}, respectively. In section~\ref{sec:kubernetes-integration} we discuss Kubernetes integration scenarios. Finally, we conclude this paper in section~\ref{sec:conclusion}. 

\section{Motivation}
\label{sec:motivation}

Building software in a reproducible fashion requires absolute control over either the build environment or the build systems involved. The responsibility to correctly specify and verify dependencies, such as which library to link against, is placed on the user. Ensuring for example consistent use of threaded vs non-threaded libraries becomes crucial for the code to execute correctly. Failure to ensure this consistency can lead to runtime errors that are difficult to track, or may even return adulterated results, despite integration testing. 

Correctly applied, containers provide a stable and controlled environment for wrapping software or even entire scientific software stacks into a portable unit. Packaging these portable units in a standardized way makes it possible to write workflows with dependencies on specific containers, rather than specific execution environments. This is in particular exploited by the bioinformatics \cite{wratten2021reproducible} and data science \cite{nust2020ten} communities, which use multiple tools with sometimes competing build and runtime environment requirements in complex data processing pipelines.

Thus, within an \gls{HPC} environment, containers aid in solving the problems of package build systems by: 

\begin{itemize}
    \item enforcing a code-based approach to the build environment and the software packaged by it via the use of container specs such as Dockerfiles or Singularity definitions,
    \item controlling the build environment such that there is only one library variant available the software can only build against it or fail at the linker step,
    \item allowing more lenient package build systems, as often found in scientific software,
    \item and bundling multiple packages into one consistent directory tree, made relocatable with the help of the container engine.
\end{itemize}

As applications are not necessarily linked statically and may need additional support files such as parameter sets, configurations, model data, etc. most applications must be run within the container they were built in or, ideally, within a pared-down version of it. These applications are fully decoupled from the system libraries on the host operating system, increasing portability. The clear advantage is that the host operating system can be updated independently, reducing the system administration burden. The drawback includes the containers not profiting from security, bugfix, or performance updates performed on the host operating system. This mandates the use of Continuous Integration/Continuous Delivery (CI/CD) systems for container update automation, and a service or registry for management and sharing. An efficient formulation of regression tests can for example be done with a software package like ReFrame \cite{ReFrame2023}.

An in-depth analysis of the possible container engines, registries, and orchestration tools aids in the selection of the most fitting tools to integrate into a full \gls{HPC} adaptive containerization architecture based on site requirements. 

\section{Concepts}
\label{sec:concepts}

The containerization stack consists of multiple components on multiple levels which must be defined. To facilitate the discussion of adaptive containerization, we must further define the set of requirements posed by \gls{HPC} sites to ensure security and efficiency for all its users. 

\subsection{Terminology}

At the lowest level is the \emph{image}, an immutable file composed of source code, libraries, and dependencies necessary for an application to run. A container image is not a container and cannot be executed. To run an image, it must be unpacked into a container - a runtime instantiation usually isolated using Linux cgroups. 

Container images can be stored in a \emph{registry} that is used to manage, push, and pull images. Registries store container \emph{repositories}, collections of related images that may have the same name and are differentiated by their tags or alphanumeric identifiers. 

The data inside a container image is usually organized in layers. A \emph{layer} captures changes in the filesystem compared to the previous layer, and is identified by a hash calculated from the data in that layer. Layer deduplication can be employed in registries and locally based on equal hashes (\emph{content-addressable storage}).

At the highest level, we differentiate between the container \emph{engines} and the container \emph{runtime interface} (CRI). Container \emph{engines} include Docker, Podman, Sarus, and many others. They permit the user to make requests regarding container images via a user-facing component. These requests may include image pulls from a registry, signature verification, unpacking of bundles, and ascertaining the availability of required system components. The engine is not a CRI, but is responsible for calling the container runtime. The container \emph{runtime} is a lower-level component that handles image and process management. The runtime sets up the \gls{UserNS}, thus starting the container process. The most popular container runtimes include \pkg{runc} and \pkg{crun}.

The interoperability between the various engines and runtimes is standardized under the \gls{OCI}. The \gls{OCI} defines a standard container image format, the container runtime specification, and a reference runtime implementation \pkg{runc}, which was split off from Docker. \gls{OCI}-compatible runtimes have \gls{OCI} Runtime Specification compliant CLI options, configuration interface and execute \gls{OCI} hooks, a mechanism that defines entry points to inject code to be run at various phases of the container lifetime. 

For building container images, as well as use cases which require more isolation based on virtualization technology, an extra set of tools and terminology is needed. Since we consider these topics outside of the scope of this paper, we hereby refer to the respective literature \cite{bentalebContainerizationTechnologiesTaxonomies2022,priedhorskyMinimizingPrivilegeBuilding2021}.

\subsection{HPC Requirements}
\label{sec:hpc-requirements}

Container technologies were initially created to run services, hence the assumption that containers are started by users with root or root-like permissions. Privilege escalation on \gls{HPC} systems poses a security issue, making alternative container execution models such as rootless a requirement. \emph{Rootless} refers to the ability to start containers as an unprivileged user and avoid running binaries with root permissions in the default or initial namespace in which every process started on a Linux system runs. This execution model provides users the capability for privilege escalation only within the container, and allows access to syscalls affecting the runtime environment, while denying access to resources on the host system to which the user does not have permissions.

While the strict adherence to rootless container execution poses one of the main challenges of deploying container technologies on \gls{HPC} systems, additional requirements differing from those of non-\gls{HPC} systems exist.

The advantages of container technologies are achieved by restricting the container process to a subset of resources and thus introducing an interface between the host operating system and the container execution environment. This interface executes a change of the filesystem root via \emph{chroot} or \emph{pivot\_root}; separates filesystem mounts, system IDs (uid, gid), process IDs, networks; etc. These restrictions in turn enable the runtime portability of containers.

Container or process isolation enables the running of multiple containers concurrently on the same hardware. Workloads on \gls{HPC} systems are often spread across multiple nodes. As compute nodes are allocated exclusively, strict container isolation may introduce performance penalties due to increased OS overhead. Note that this penalty is in flux as the use of exclusive node allocation has been shifting for high-density systems with more than two GPUs per node, with enough cores per node, or for workflows that require direct communication between applications in different containers by means of e.g. shared memory. 

Beyond the introduced overhead, strict container isolation may break access to \gls{HPC} hardware such as interconnects or accelerators. Additionally, it may break software with custom inter-node communication such as e.g. workflow-based payloads that require non-MPI communication. 

Filesystem access poses another set of issues. Container processes may require access to specific files on the host system, including device libraries, configuration files, license restricted files, etc. When loading host libraries for device drivers, communication, etc., ABI compatibility with the container applications and libraries must be ensured. Failure to do so may lead to errors which are hard to detect and may possibly affect scientific results.
This may extend beyond the requirements of the directly involved libraries: if a host library imported into the container requires a newer version of \texttt{glibc} than present within the container it will fail. Overriding most of the container libraries with the host supplied ones may equally introduce compatibility issues.

Interoperability with data on existing shared filesystems is important, whereas large scale container deployments in the cloud often avoid shared data or POSIX filesystems altogether. A container image contains many small files which may be loaded from shared storage from many compute nodes and that put strain on the cluster filesystem, slowing down startup time or even execution. This can also affect other users of the shared storage.

Software within containers must be optimized for a target architecture, breaking portability. While tracking dependencies for security issues across the layers of a container is crucial for cloud service containers, it is less important for \gls{HPC} containers as the applications usually do not offer a continually run service which can be exploited from arbitrary locations, thus adding a level of isolation. Nonetheless, there are attack scenarios which may require scanning images as due diligence.

Finally, on \gls{HPC} systems malicious users must be assumed, and privilege escalation avoided. Possible attack vectors if a user is granted elevated privilege include the injection of malicious data in filesystem images to exploit kernel block device driver security bugs. Spinning up a daemon on each compute node to control what is most often a single container process is wasteful and may introduce extra jitter, and increases the attack surface in the process.

The various \gls{HPC} container technologies implemented to date share similar solutions to the above stated requirements:

\begin{itemize}
    \item \gls{HPC} container solutions do not require running daemons to start containers, and especially root or root-like daemons must be avoided. Tasks such as registry interaction may use daemons. A monitoring process to give enhanced control over the running process may still be involved, but must run as the same user starting the process.
    \item Container isolation is weakened, resulting in a setup which offers more isolation than a simple \texttt{chroot}, but less than full container isolation. A user \texttt{namespace} is created to obtain extra capabilities within the container, which are in turn used to set up separate mounts invisible to everyone beyond the real root of the host system. This enables overlaying parts of the host OS into the container environment, allowing the container processes to access host OS libraries or devices.
    \item Unused isolations such as \texttt{network} or \texttt{IPC} namespaces are not set up to reduce complexity and attack surface, or because they may interfere with \gls{HPC} applications. \gls{HPC} workload managers such as \pkg{Slurm} or the PBS family are expected to be configured for exclusive node allocation. 
    \item Host configurations are copied into the container environment, modifying it in the process. User namespacing is limited to a single user to ensure files created by processes in the container have the UID/GID of the user launching the job.
    \item Container filesystems are (re-)packaged as single-file images to avoid small-file load and latency, potentially providing a speedup against traditional application execution by trading memory and CPU (decompression) for disk IO.
    \item When a kernel driver for compressed filesystem image reading is used, care is taken that the user can never directly provide such a compressed image. Caching of such filesystem images may thus require a separate service or to be run as root, or a userspace filesystem driver must be used to avoid exposing the kernel to user generated filesystem images.
\end{itemize}

The containerization technology space is a quickly developing one. Many tools exist that enable the building and running of containers. As such, the following analysis makes no claims to completeness as new technologies are developed constantly. Nevertheless, great care has been taken to analyze the requirements and the solutions given by containerization technologies with a focus on \gls{HPC}.

\section{Container Engine Comparison}
\label{sec:container-engine-comparison}

This section contains a set of tables giving an overview of features we identified as important for \gls{HPC} Containers. The chosen criteria are used to evaluate the most prominent (\gls{HPC}) container solutions which may be deployed without extra management services such as Kubernetes, Docker Swarm, Apache Mesos, OpenShift, Rancher, or Nomad. Additionally, we include Docker as a baseline comparison and for the sake of completeness. Tables \ref{tab:container-solutions}, \ref{tab:container-solutions-formats} and \ref{tab:container-solutions-integration} contain the summarized comparison for the following discussion.

\begin{table*}
\begin{subtable}[h]{\textwidth}
\footnotesize
\centering
\begin{tabular}{@{}llllll@{}} \toprule
    \textbf{Engine}
        & \textbf{Version}
        & \textbf{Champion}
        & \textbf{Affiliation}
        & \textbf{Runtime}
        & \textbf{Implem. Language}
        \\
    \midrule
    \textbf{Docker}        & v24.0.5 (Jul. 24, 2023)     & Docker     & Docker           & \pkg{runc}/\pkg{crun}        & Go      \\
    \textbf{Podman}~\cite{Podman} & v4.6.1 (Aug. 10, 2023)      & RedHat/IBM & Kubernetes       & \pkg{crun}/\pkg{runc}/Crio-O & Go      \\
    \textbf{Podman-HPC}~\cite{Podman-HPC} & v1.0.2 (Jun. 15, 2023)      & NERSC      & -                & \pkg{crun}/\pkg{runc}/Crio-O & Python, C \\
    \textbf{Shifter}~\cite{jacobsenContainThisUnleashing} & Git 0784ae5 (Oct. 22, 2022) & NERSC      & -                & Shifter          & C       \\
    \textbf{Sarus}~\cite{benedicicSarusHighlyScalable2019} & v1.6.0 (May 5, 2023)        & CSCS       & -                & \pkg{runc}/\pkg{crun}        & C++     \\
    \textbf{Charliecloud}~\cite{priedhorskyCharliecloudUnprivilegedContainers2017} & v0.33 (Jun. 9, 2023)        & LANL       & -                & Charliecloud     & C       \\
    \textbf{Apptainer}~\cite{Apptainer} & v1.2.2 (Jul. 27, 2023)      & LLNL, CIQ  & Linux Foundation & \pkg{runc}/\pkg{crun}        & Go      \\
    \textbf{SingularityCE}~\cite{SingularityCE} & v3.11.4 (Jun. 22, 2023)     & Sylabs     & -                & \pkg{crun}/\pkg{runc}        & Go      \\
    \textbf{ENROOT}~\cite{ENROOT} & v3.4.1 (Feb. 8, 2023)       & Nvidia     & Nvidia           & enroot           & C, Bash \\
    \bottomrule
\end{tabular}
\end{subtable}
\newline
\vspace*{2mm}
\newline
\begin{subtable}[h]{\textwidth}
\footnotesize
\centering
\begin{tabular}{@{}llllll@{}} \toprule
    \textbf{Engine}
        & \textbf{Rootless}
        & \textbf{Rootless-FS}
        & \textbf{Container Monitor}
        & \multicolumn{2}{c}{\textbf{\gls{OCI} Support}}
        \\
    
        & 
        & 
        &
        & \textbf{Hooks}
        & \textbf{Container}
        \\
    \midrule
    \textbf{Docker}
        & \gls{UserNS}
        & fuse-overlayfs
        & per-machine (dockerd)
        & yes
        & yes
        \\
    \textbf{Podman}
        & \gls{UserNS}
        & fuse-overlayfs
        & per-container (conmon)
        & yes
        & yes
        \\
    \textbf{Podman-HPC}
        & \gls{UserNS}
        & SquashFUSE + fuse-overlayfs
        & per-container (conmon)
        & yes
        & yes
        \\
    \textbf{Shifter}
        & \gls{UserNS}
        & suid
        & no
        & no
        & yes (partial)
        \\
    \textbf{Sarus}
        & \gls{UserNS}
        & suid
        & no
        & yes
        & yes (partial)
        \\
    \textbf{Charliecloud}
        & \gls{UserNS}
        & Dir, SquashFUSE
        & no
        & no
        & yes (partial)
        \\
    \textbf{Apptainer}
        & \gls{UserNS}, fakeroot\cite{dykstraApptainerSetuid2022}
        & suid, fakeroot, (SquashFUSE)
        & per-container (conmon)
        & yes (manually, requires root)
        & yes (partial)
        \\
    \textbf{SingularityCE}
        & \gls{UserNS}, fakeroot
        & suid, fakeroot, SquashFUSE
        & per-container (conmon)
        & yes (manually, requires root)
        & yes (partial)
        \\
    \textbf{ENROOT}
        & \gls{UserNS}
        & Dir
        & no
        & no
        & yes (partial)
        \\
    \bottomrule
\end{tabular}
\end{subtable}
\caption{Overview of container engines with supported \emph{rootless}-techniques and \gls{OCI} compatibility.}
\label{tab:container-solutions}
\end{table*}

\subsection{Discussion}

In the following, some of the chosen evaluation criteria are defined, examined, and discussed. Note that where no distinction between Apptainer, SingularityCE, and SingularityPro is required, we refer to them collectively as \textit{Singularity}.

\subsubsection{Champion and Affiliation}

In the past, project affiliation was less significant to \gls{HPC} sites, as they were more isolated and compatibility with bleeding edge technology played a different role. While all listed projects are Open Source Software (OSS), the entities behind them can retain significant influence. This influence may range from limiting features to paying customers, affecting the development trajectory, up to complete project defunding.

% An example of how affiliation may influence the development of an application is the planned\todo{check if still in planning or done} site-wide change in container technology use at the National Energy Research Scientific Computing Center (NERSC) to podman. While Shifter was deployed on their flagship Perlmutter, and will likely be maintained for the planned runtime, it must be assumed that Perlmutter will follow the site-wide change at some point and replace Shifter in favor of podman. \todo{I'm not sure if this is a good example, both list NERSC as champion}

An example of this dynamic is the interaction between Apptainer and SingularityCE. The company behind the Singularity platform, which diverged into Apptainer and SingularityCE, has released SingularityPro with additional features (e.g. \gls{SBOM}) and support contracts, as well as their own maintained registry platform. While it can be advantageous to externalize some of the costs of maintaining base images, it can also be seen as an attempt at the platformization of the \gls{HPC} container space. Additionally, despite Apptainer incorporating changes made in SingularityCE, a quick comparison shows differences between them, with e.g. Apptainer using \texttt{runc} and SingularityCE using \texttt{crun} as their default runtimes. As Sylabs develops its business, it is safe to assume that more of the advanced features will be incorporated into SingularityPro and may not be open sourced.

\subsubsection{Rootless container/Rootless-FS implementation}
\label{sec:rootless-container}

The core principal behind rootlessness is the use of \texttt{pivot\_root} instead of the classical \texttt{chroot} to provide a new root to the processes started in the container. A user gains the capability to \texttt{pivot\_root} when in their own \gls{UserNS}. Despite the user being able to assume \texttt{UID 0} inside of this new namespace, it does not permit mounting block devices or files acting as such via kernel drivers, since kernel drivers are not hardened against maliciously crafted block-device data. Therefore, a SquashFS image can only be mounted by either a setuid-root binary prior to entering the namespace, via a FUSE driver as the FUSE user-kernel interface can be assumed to be audited, or not at all, instead unpacking an image to a directory. When using the setuid-root approach, care must be taken to not only secure the binary and the image itself (e.g. on transparent conversion from an \gls{OCI} image layer to a SquashFS image at runtime, the resulting image must not be user-writeable), but also to ensure that the user is unable to manipulate it while being mounted, nor inject their own image directly.

One approach that works around the limitations imposed by a shared cluster filesystem  is extracting an image to a temporary, node-local storage location. This avoids the need for either user- or kernel-space filesystem drivers, thus reducing the memory and CPU overhead generated by the intermediate image.
% the things above are at odds with on-the-fly decryption:
% Further, it allows an approach which utilizes images as an intermediate encryption layer to enable on-the-fly decryption scenarios.

An alternative to the namespace-based rootless mechanisms are the \texttt{fakeroot} approaches: an \texttt{LD\_PRELOAD} variant, in which a library intercepting relevant system calls is loaded prior to any executable; or a variant based on the \texttt{ptrace} system call, allowing to intercept the system calls of another process. A limitation of the first approach is that it fails with static binaries, and for the second that it introduces a significant performance penalty and the user requires access to the \texttt{CAP\_SYS\_PTRACE} capability.

We note that benchmarks comparing SquashFUSE and the in-kernel SquashFS show a magnitude lower IOPS for random access and a much higher latency\cite{stoppelsSquashfsmount2023}. For many compiled codes this will only be noticeable on start and when loading bundled parameter data, while for projects based on interpreted languages like Python which consist of many small files it will have a more noticeable effect. A similar situation can arise with a FUSE-based OverlayFS implementation, where heavy I/O must absorbed by the CPU.

\begin{table*}
\footnotesize
\centering
\begin{tabular}{@{}lllllll@{}} \toprule
    \textbf{Engine} &
        \textbf{\begin{tabular}[c]{@{}l@{}}Transparent \\ Format \\ Conversion\end{tabular}} &
        \textbf{\begin{tabular}[c]{@{}l@{}}(Transparent) \\ Native Container \\ Format Caching\end{tabular}} &
        \textbf{\begin{tabular}[c]{@{}l@{}}Native \\ Format \\ Sharing\end{tabular}} &
        \textbf{Namespacing on Execution} &
        \textbf{\begin{tabular}[c]{@{}l@{}}Signature Verification \\ Support\end{tabular}} &
        \textbf{\begin{tabular}[c]{@{}l@{}}Encrypted Container \\ Support\end{tabular}} \\ \midrule
    \textbf{Docker}                   & -   & -   & -   & full                               & Notary                        & no, extensions available          \\
    \textbf{Podman}                   & -   & -   & -   & full                               & GPG, sigstore                 & yes                               \\
    \textbf{Podman-HPC}               & yes & yes & no  & full/user and mount NS             & GPG, sigstore                 & yes                               \\
    \textbf{Shifter}                  & yes & yes & no  & user and mount NS                  & -                             & no                                \\
    \textbf{Sarus}                    & yes & yes & yes & user and mount NS                  & -                             & no                                \\
    \textbf{Charliecloud}             & no  & -   & no  & user and mount NS                  & -                             & no                                \\
    \textbf{Apptainer}                & yes & yes & yes & user and mount NS, possibly others & GPG (SIF containers) & yes (SIF only, via kernel driver) \\
    \textbf{SingularityCE}            & yes & yes & yes & user and mount NS, possibly others & GPG (SIF containers) & yes (SIF only, via kernel driver) \\
    \textbf{ENROOT}                   & no  & -   & no  & user and mount NS                  & -                             & no \\
    \bottomrule
\end{tabular}
\caption{Continuation of table~\ref{tab:container-solutions} with focus on supported image formats and features.}
\label{tab:container-solutions-formats}
\end{table*}

\subsubsection{\gls{OCI} Container and Hook support}

Support for hooks becomes important if extensions such as for additional image modifications or accelerator enablement are required. The \gls{OCI} hooks specification, which is part of the \gls{OCI} runtime spec, provides a vendor-independent way of installing and running such hooks at defined points in the lifetime of a container without the need to modify the runtime itself. Most solutions either provide direct support for \gls{OCI} hooks (in particular when relying on a mainstream runtime like \pkg{runc} or \pkg{crun}), or a custom hook framework (see plugins in Apptainer). Some container solutions, such as e.g. Shifter, rely on parts of their software being written in a scripting language, therefore making them easily extendable and more flexible, but requiring adaptions of the extensions when the base software is updated.

For \gls{OCI} compatibility, it is important to note that \gls{HPC} container solutions, as well as the solutions adapted for \gls{HPC} like Podman-HPC, break some of the features a container expects to be present. The most obvious of these are the lack of an isolated network namespace which permits the binding of services to arbitrary ports, or the availability of different user IDs, as often only a single one mapped directly to the original user ID is made available. Thus vanilla containers may have to be repackaged or modified to run on an \gls{HPC} container system.

\subsubsection{Container Format, Conversion, Caching and Registry}

The Singularity Definition file \texttt{.def} is similar to RPM specs, and all commands to build the container can be placed in a single section, as layering is not available in the flat Singularity Image Format \gls{SIF}. \gls{SIF} integrates writable overlay data, which may be useful to bundle either models or output data with the code using or generating it. In Dockerfiles, on the other hand, manually grouping commands into layers poses an important concept to allow incremental container builds, updates, and deployments.
% For \gls{OCI} containers, data handling is outside of the specification, even though separate initiatives exist in the cloud computing container space to integrate machine learning models in the container registries, like the open-source model registry \texttt{ormb}.
% ^^^ not entirely sure anymore where I wanted to go with the "outside of the spec" argument, using registries to store arbitrary data is now standardised in parts by Artifacts and driven by the ORAS project
It has to be noted that in a setup integrating Podman and Singularity, since Podman is capable of running \gls{SIF} containers, Singularity may be needed solely to build the container with Podman capable of launching it. 

In non-\gls{HPC} containers, the \gls{OCI} filesystem bundle consists of multiple layers, with each layer tracking a change to previous filesystem layers. These layers are mounted via a union mount filesystem approach - usually the Linux based OverlayFS driver - into a consistent filesystem view with only a new upper layer being writable. Running an \gls{OCI} container in an \gls{HPC} setting thus requires all layers to be present on a shared storage. This overlay mount capability may not be enabled on the compute nodes, or may require root privileges depending on the kernel version.

\gls{HPC} cluster filesystems, or any shared filesystems, are known for not scaling well in cases of random access with many small files, as can be seen with e.g. Python or other interpreters.
Even for more static workloads, there may be additional load to the shared filesystem on startup. This load is caused by the containers' base libraries, which have to be loaded into memory, and which may load additional service files that would otherwise already be in memory due to the host operating system requiring them. For example, \pkg*{libc} will have to load \fpath{/etc/nsswitch.conf} to determine the source of UIDs/GIDs, and from there the respective configuration files; or system libraries may have to load localization specifications, etc. As established earlier, one solution to work around these limitations is to flatten the \gls{OCI} bundle either to a node-local directory, or to a filesystem image on a shared storage. This conversion can happen either automatically or explicitly. In the automatic case, we want this converted image to be cached to avoid repeated conversion costs (storage and time), and possibly share it between different users.

If an in-kernel driver is used to mount a filesystem image, care must be taken that the user can neither provide nor manipulate the image directly, instead using a userspace driver (SquashFUSE) to mount it, as pointed out in sections \ref{sec:hpc-requirements} and \ref{sec:rootless-container}. When an in-kernel driver is used, there must be a setuid-root binary doing the conversion, caching, and sharing between multiple users, as in e.g. Sarus. It must be noted that an OverlayFS mount does not suffer from the same risks as a SquashFS mount, since the OverlayFS does not access raw block device data, but acts on the mounted filesystem instead. Both Docker and Podman switched to a FUSE-based OverlayFS \pkg*{fuse-overlayfs} driver.

Since sharing container images - irrespective of format - via filesystem has several challenges (e.g. managing access via POSIX ACLs or extended ACLs, deduplication, locking), a better approach may be a separate service which can ensure the required access semantics. Sarus and Singularity support sharing their respective \emph{native} \gls{HPC} container format, while other solutions do not and sharing always happens prior to conversion to the format, or by manual setup directly between users.

\subsubsection{Container Signing and Encryption}

While digital signing does not prevent the spread of malware, or protect from well-funded malicious actors, it can help uncover basic attacks in the form of name squatting, breaches of security, and make tracing software provenance possible.
Docker and Podman, the industry solutions at the forefront of digital signing, implement different solutions:
Docker uses the Notary (v2) tool, while Podman provides similar functionality via GPG signature attachments.
Apptainer has built its signing solution on PGP as well, although only for its own \gls{SIF} container, meaning that signatures for imported \gls{OCI} containers are not verified.
\pkg{sigstore}, with \pkg{cosign} being the implementation for containers, is an independent approach which can be used for general software signing support, and for supporting \gls{SBOM} or lists of all the open-source and third-party components present.
% this might not be 100% correct: sigstore is the broader ecosystem while cosign is just the signing part
Podman added direct support for \pkg{cosign}, but it currently requires extra setup steps.

Running code on or transferring data to external machines always requires a certain amount of trust.
Container encryption can limit data access times and access privileges.
When combined with secure enclaves, it can make it difficult to access data despite having total control over the hardware the data is stored on.
This permits the deployment of workloads which previously had to be run on isolated systems on a supercomputer.
Since this feature is still under development for most solutions, we tracked only the simplest form of support: does the runtime, resp. engine, support decryption of encrypted containers.

\subsubsection{\gls{HPC}-specific extensions - GPU and Accelerator Enablement, Library Hookup, WLM Integration}

As mentioned in section~\ref{sec:hpc-requirements}, \gls{HPC} container technologies require additional steps at container startup. This includes granting containers access to host libraries like optimized scientific libraries, device libraries for accelerators, communication, etc. Host library access can be enabled by bind-mounting host directories into the container namespace, providing extra device nodes, or granting extra capabilities to the user process. Most \gls{HPC} container implementations have solutions built-in for the GPUs of the most common vendors. Other accelerators must be added via hooks or plugins. When a container gains access to host libraries, it requires a matching ABI, as a mismatch may introduce subtle errors.
Some solutions like Sarus therefore contain explicit ABI compatibility checks on the libraries.

While it is possible for all container solutions to be invoked explicitly via batch scripts, proper \gls{WLM} integration may be required. The \gls{WLM} controls device access rights, which must be passed along to the container engine, and may restrict the capabilities available to the user (like cgroups). A more transparent container execution may be desirable in order to lower the container entrance barrier for users and avoid common mistakes when running containers.

As containers introduce an additional layer of indirection, some workflows using interactive access may be broken or require additional steps. This problem could be alleviated via proper \gls{WLM} integration.
Other use cases involving profilers and debuggers may instead require approaches specifically tailored to container usage in \gls{HPC}.

\subsubsection{Module System Integration}

While it is possible to write a wrapper script to transparently start a container in which to run an application, doing so may have unexpected side-effects for the user and require additional work. With the exception of the Singularity Registry HPC (\pkg{shpc}), none of the other projects offer affiliated solutions to automatically integrate containers as modules. Despite \pkg{shpc} originating in the Singularity ecosystem, it officially supports other container solutions like \pkg{Podman}, although they may require additional configuration in the form of wrapper scripts.

\begin{table*}
\begin{subtable}[h]{\textwidth}
\footnotesize
\centering
\begin{tabular}{llllll} \toprule
    \textbf{Engine}
        & \textbf{GPU-Enablement} & \textbf{Accelerator Support}
        & \textbf{OS/MPI Library Hookup}
        & \textbf{WLM Integration}
        & \textbf{Contains Build Tool}
        \\
    \midrule
    \textbf{Docker} & via \gls{OCI} hooks & via \gls{OCI} hooks & via \gls{OCI} hooks & no & yes \\
    \textbf{Podman} & via \gls{OCI} hooks & via \gls{OCI} hooks & via \gls{OCI} hooks & no & yes \\
    \textbf{Podman-HPC} & yes & via \gls{OCI} hooks or patch & yes & no & yes  \\
    \textbf{Shifter} & no & no & for MPICH & yes / SPANK plugin & no  \\
    \textbf{Sarus} & yes & via \gls{OCI} hooks & yes & partially via \gls{OCI} hooks & no \\
    \textbf{Charliecloud} & manually & manually & manually & no (no SPANK plugin release) & no  \\
    \textbf{Apptainer} & yes & no & manually & no & yes \\
    \textbf{SingularityCE} & yes & no & manually & no & yes \\
    \textbf{ENROOT} & yes, Nvidia only & via custom hooks & via custom hooks & yes / SPANK plugin & no  \\
    \bottomrule
\end{tabular}
\end{subtable}
\newline
\vspace*{2mm}
\newline
\begin{subtable}[h]{\textwidth}
\footnotesize
\centering
\begin{tabular}{llllll}
\toprule
    \textbf{Engine}
        & \textbf{Module System Integration}
        & \multicolumn{3}{l}{\textbf{Documentation}}
        & \textbf{\#\,Contributors}
        \\
        & 
        & \textbf{User}
        & \textbf{Admin}
        & \textbf{Source}
        & 
        \\
    \midrule
        \textbf{Docker}        & via shpc            & +++ & +   & +     & 486  \\
        \textbf{Podman}        & via shpc            & +   & N/A & ++    & 461  \\
        \textbf{Podman-HPC}    & (via shpc)          & N/A & N/A & ( + ) & 3    \\
        \textbf{Shifter}       & no (shpc announced) & +   & +   & ++    & 17   \\
        \textbf{Sarus}         & no (shpc announced) & ++  & ++  & +     & 6    \\
        \textbf{Charliecloud}  & no                  & +++ & +   & ++    & 31   \\
        \textbf{Apptainer}     & via shpc            & ++  & +   & +     & 148  \\
        \textbf{SingularityCE} & via shpc            & ++  & N/A & +     & 130  \\
        \textbf{ENROOT}        & no                  & N/A & N/A & +     & 9    \\
    \bottomrule
\end{tabular}
\end{subtable}
\caption{Summary of supported integrations for different container solutions and community analysis.}
\label{tab:container-solutions-integration}
\end{table*}

\subsubsection{Documentation}

Grading of the software documentation for each engine is based on three elements - the availability and length of the documentation, the breadth and depth of the topics covered, as well as the clarity of the text. While the grading must by definition be subjective we hope to somewhat standardize it with the given criteria. Where no documentation (e.g. Podman-HPC) or insufficient documentation was available, the corresponding slot has been marked with N/A. In all other cases, a scale ranging from \texttt{+} for minimal documentation available to \texttt{+++} for extensive and well-organized documentation was used.

\subsubsection{State of Source Code, Contributors, Community}

The number of contributors and/or size of the community may serve as an indicator of the future development of a project: a small contributor base originating mostly from a single entity can change the direction of a project drastically, whereas defunding of the project within that entity can bring it to a sudden stop. We see this risk in particular for the Shifter, Sarus, Charliecloud, and Enroot projects, but also for the Podman-HPC project, which is at the moment at an incubator stage. For the latter it is possible that required features will be directly ported to \pkg{Podman}.
For operations this could mean that system administrators have to take over maintenance at a lower level than initially anticipated. This must be accounted for in the risk assessment.

Shifter and Podman-HPC are both developed at NERSC. Based on presentations by NERSC \cite{fultonContainersHPCShifter2022} it has to be assumed that Shifter will be replaced by a solution based on \pkg{Podman}, with the Podman-HPC development serving as a testing ground (intercepting layer unpacking to generate a filesystem image; using \gls{OCI} hooks to setup devices and device libraries; integrating with the WLM).

It must be stressed that the number of contributors does not paint the whole picture. In particular, while SingularityCE has fewer contributors than Apptainer, the activity in the SingularityCE repository as measured by the number of added/deleted lines for November 2022 was twice that of the activity of the Apptainer repository.

\subsection{Summary}

With regard to \gls{HPC} requirements, the container technology selection space is essentially tripartite: cloud industry container tools like Docker and Podman, Singularity with their own image format, and projects trying to integrate containers without deviating too much from the cloud industry. Industry tools have the advantage of an immense user community with a lot of resources, while \gls{HPC}-centric tools may be more accommodating to the domain scientists, such as by using the simpler structure of the Singularity container specification files. 

It is difficult to make specific assessments regarding the future-proofness of any single solution.
The \gls{HPC} space has been trending towards using Singularity in previous years.
Big cloud providers are pushing into the \gls{HPC} segment and have started to support \gls{SIF} containers in their registries to ease the onboarding of scientific compute customers on their platforms.
Newer tools like Podman support running containers from \gls{SIF}, easing the shift from Singularity in case of adoption of Podman for existing deployments.
While this Podman support does not indicate that \gls{SIF} will become a standard used outside of scientific computing, it shows that there is support for it and that there are implementations beyond Singularity capable of executing it.
\gls{HPC} initiatives like Autamus provide both kinds of image formats, anticipating the needs of Singularity alternatives relying solely on \gls{OCI} containers.
Together with NERSC gravitating towards Podman (with the help of a small wrapper and support from RedHat), Sylabs improving the support for \gls{OCI} containers for SingularityCE 4, and Apptainer gaining support for building from Dockerfiles, the longterm prospects of \gls{SIF} become less certain.

\section{Container Registry and CI/CD comparison}
\label{sec:registry-comparison}

An internally deployed Container Registry may be required to maintain containers and to act as an intermediary instead of SSH transfers.
Here we compare existing solutions deployable either on-premise or with CI/CD integration.
Some cloud platforms like OpenShift provide built-in registries. As evaluating such a cloud platform is out-of-scope for this document, these registries have been omitted from the comparison. We refer to tables \ref{tab:registries} and \ref{tab:registries-integration} for the list of the included container registry projects and the criteria discussed in the following section.

\subsection{Discussion}

\begin{table*}
\begin{subtable}[h]{\textwidth}
\footnotesize
\centering
\begin{tabular}{@{}llllll@{}} \toprule
    \textbf{Registry}
        & \textbf{Version}
        & \textbf{Champion}
        & \textbf{Affiliation}
        & \textbf{Focus}
        & \textbf{Protocol}
        \\
    \midrule
    \textbf{Quay}~\cite{Quay}
        & v3.8.10 (Dec. 6 2022)
        & RedHat/IBM
        & -
        & Registry
        & \gls{OCI} v2
        \\
    \textbf{Harbor}~\cite{Harbor}
        & v2.8.3 (Jul. 28, 2023)
        & VMWare
        & CNCF
        & Registry
        & \gls{OCI} v2
        \\
    \textbf{GitLab}
        & v16.2 (Jul. 22, 2023)
        & GitLab
        & -
        & Git hosting, CI/CD
        & \gls{OCI} v2
        \\
    \textbf{Gitea}
        & v1.20.2 (Jul. 29, 2023)
        & (OSS community)
        & -
        & Git hosting, CI/CD
        & \gls{OCI} v2
        \\
    \textbf{shpc}~\cite{shpc}
        & v2.1.0 (Apr. 6, 2023)
        & vsoch
        & LLNL
        & Registry
        & Library API
        \\
    \textbf{Hinkskalle}~\cite{Hinkskalle}
        & v4.6.0 (Oct. 18, 2022)
        & h3kker
        & University of Vienna
        & Registry
        & Library API, \gls{OCI} v2
        \\
    \textbf{zot}~\cite{zot}
        & v1.4.3 (Nov. 30, 2022)
        & Cisco
        & CNCF
        & Registry
        & \gls{OCI} v1
        \\
    \bottomrule
\end{tabular}
\end{subtable}
\newline
\vspace*{2mm}
\newline
\begin{subtable}[h]{\textwidth}
\footnotesize
\centering
\begin{tabular}{@{}llllp{.2\linewidth}p{.2\linewidth}@{}} \toprule
    \textbf{Registry}
        & \textbf{\gls{OCI} Artifact Support}
        & \textbf{Proxying}
        & \textbf{Repl./Mirroring}
        & \textbf{Storage Support}
        & \textbf{Authentication Providers}
        \\
    \midrule
    \textbf{Quay}
        & Helm charts, cosign, zstd
        & yes / auto
        & yes (pull)
        & FS, S3, GCS, Swift, Ceph
        & internal, LDAP, Keystone, OIDC, Google, GitHub
        \\
    \textbf{Harbor}
        & Helm charts, cosign, user-def.
        & yes / auto
        & yes (push + pull)
        & FS, Azure, GCS, S3, Swift, OSS
        & internal, LDAP, UAA, OIDC
        \\
    \textbf{GitLab}
        & no, separate pkg registries
        & yes / manual
        & no
        & FS, Azure, GCS, S3, Swift, OSS
        & LDAP
        \\
    \textbf{Gitea}
        & Helm, separate pkg registries
        & no
        & no
        & FS, Minio/S3
        & internal, LDAP, PAM, Kerberos
        \\
    \textbf{shpc}
        & -
        & no
        & manual (Globus)
        & Minio, GCS, S3
        & LDAP, PAM, SAML
        \\
    \textbf{Hinkskalle}
        & no
        & no
        & no
        & FS
        & LDAP
        \\
    \textbf{zot}
        & Helm charts, cosign, notation
        & no
        & yes (pull)
        & FS, S3
        & internal, LDAP
        \\
    \bottomrule
\end{tabular}
\end{subtable}
\caption{List of common container registries with their respective featureset.}
\label{tab:registries}
\end{table*}

\subsubsection{Champion and Affiliation}

While the landscape of container solutions and build tools is diversified, the same can not be said for registry services. Since registries can be provided as-a-Service, most registries have a company sponsored development.
This increases the risk of the product taking either new directions or having the development model or license switched.
The project documentation often reflects this. The Project Quay documentation often refers directly to Red Hat services.

Even though Apptainer or SingularityCE can push and pull standardized \gls{OCI} containers, a separate \gls{SIF} compatible registry adhering to the Library API standard may be advisable for improved integration (signing, avoiding repackaging, preserving metadata, etc.).
Besides the as-a-Service registry provided by Sylabs, which is not part of this comparison, the available Library API registries are maintained by single developers and are likely to receive less scrutiny than the \gls{OCI} registries.

\subsubsection{Focus}

Several CI/CD solutions offer package registries with the intention to directly host build artifacts. While they explicitly offer container registries, their feature sets are limited and they may not be suitable for extensions like the ability to store, verify, and display signatures, or accept artifacts other than containers. A broad support of the \gls{OCI} standard is crucial for the development of the Adaptive Containerization feature as it could build on user-defined \gls{OCI} artifacts.

\subsubsection{Proxying Capabilities and Mirroring}

The most popular public \gls{OCI} registry DockerHub introduced rate limiting.
Any site with a small number of public IP addresses for a large number of clients is quickly affected by this.
While layers originating from public containers will be cached on an internal registry, the upstream registries may still be regularly queried when containers are (re)built, or when public containers are used on a system where at least a part of the nodes have unfettered internet access.
One way to work around such a limitation is the use of a proxy server to cache the requests.

A registry implementing proxy capabilities by means of transparently forwarding and caching requests in a namespace to an upstream registry can provide such proxy services.
The advantages over a common HTTP(S) proxy include detailed statistics about upstream registry usage, required disk space, image statistics, etc.
The mirroring capabilities can be used to either mirror hosted containers to a public registry, or to preserve remotely provided containers on the local infrastructure.
Finally, we would like to note that deploying a site-local registry with proxying capabilities possibly attached to a clusters' highspeed network can support public network access scenarios without giving login or compute nodes full internet access.

\begin{table*}
\footnotesize
\centering
\begin{tabular}{@{}llllllll@{}} \toprule
\textbf{Registry} &
  \textbf{\begin{tabular}[c]{@{}l@{}}Image \\ Squashing\end{tabular}} &
  \textbf{\begin{tabular}[c]{@{}l@{}}Image \\ Formats\end{tabular}} &
  \textbf{Multi-Tenancy} &
  \textbf{Quota} &
  \textbf{Signing} &
  \textbf{Deployment} &
  \textbf{Build Integration} \\
    \midrule
  \textbf{Quay}&
  on-demand &
  \gls{OCI} &
  yes ("Organization") &
  per-project &
  yes &
  Kubernetes Operator &
  build on Kubernetes, EC2 \\
  \textbf{Harbor} &
  no &
  \gls{OCI} &
  yes ("Project") &
  per-project &
  yes &
  Docker Compose, Helm Chart &
  via CI/CD \\
  \textbf{GitLab} &
  no &
  \gls{OCI} &
  yes ("Organization") &
  {\color[HTML]{000000} \begin{tabular}[c]{@{}l@{}}minimal solution\\ self-hosted\end{tabular}} &
  no &
  \begin{tabular}[c]{@{}l@{}}Linux packages, Helm Chart, \\ Kubernetes Operator, Docker, GET\end{tabular} &
  via CI/CD \\
  \textbf{Gitea} &
  no &
  \gls{OCI} &
  no &
  no &
  no &
  Docker Compose, Binary, Helm Chart &
  via CI/CD \\
  \textbf{shpc} &
  - &
  \gls{SIF} &
  no &
  no &
  yes &
  Docker Compose &
  build on GCC \\
  \textbf{Hinkskalle} &
  - &
  \gls{SIF}, \gls{OCI} &
  no &
  no &
  yes &
  Docker Compose &
  no \\
  \textbf{zot} &
  no &
  \gls{OCI} &
  no &
  no &
  yes &
  Docker, Helm, Podman &
  via CI/CD
  \\
    \bottomrule
\end{tabular}
\caption{Continuation of table~\ref{tab:registries}, listing supported image formats, deployment techniques and build integration.}
\label{tab:registries-integration}
\end{table*}

\subsection{Summary}

With the Library API (Singularity) registries being carried by single developers, and the CI/CD system integrated registries supporting only a subset of the possibly required features (proxying, mirroring, user-defined \gls{OCI} artifacts), the remaining candidates for an \gls{HPC}-centric container setup are Project Quay and Harbor.
Since Harbor is a Cloud Native Computing Framework project, it seems to currently enjoy broad support despite being sponsored by VMWare.
Comparing the number of contributors (around 260 for Harbor vs 60 for Quay) lends credence to this assessment.

Since \gls{SIF} images can be pushed to \gls{OCI} registries, as demonstrated by the Singularity \gls{HPC} Library where container images are hosted on either DockerHub or the Github Container Registry, there is no technical requirement to deploy a Library API registry when choosing Singularity.

\section{Kubernetes Integration Scenarios}
\label{sec:kubernetes-integration}

In previous sections, we evaluated the integration of container engines with \glspl{WLM} such as \pkg{Slurm}.
These \glspl{WLM} schedule jobs and execute batches of single execution jobs on an \gls{HPC} cluster based on predefined and flexible rules.

In a cloud system, a similar function may be performed by a cloud orchestrator - an application that can configure, manage, and coordinate multiple containers.
The best known of these is \pkg{Kubernetes}, an open source platform first developed by Google, but now maintained by the \gls{CNCF}.
Various distributions of Kubernetes exist, including \pkg{K3s} (\texttt{lightweight Kubernetes}), a fully conformant, pared down version packaged in a single binary and designed for use on e.g. Edge, IoT, Embedded, etc. devices.

The interest in \pkg{Kubernetes} integration scenarios for \gls{HPC} systems stems from the fact that domains such as bioinformatics and the data sciences have made successful use of traditional cloud resources like the \gls{GCP} for their research pipelines.
Subsequently, workflows and workflow systems have been developed which rely on Kubernetes as an interface to deploy and run the respective containers and processes.
Supercomputing facilities, meanwhile, optimized the use of \glspl{WLM} on their systems in the past decades. They must thus be able to integrate their \glspl{WLM} with orchestrators when running such \pkg{Kubernetes} workloads. This is particularly crucial in regards to the accounting of used resources, but also for optimized scheduling and the integration of newer features like preemption.

As noted in \ref{sec:hpc-requirements}, \gls{HPC} container software often provides only a subset of the isolation techniques cloud-native solutions employ, and some containers may therefore require modification before use.
Integrating \pkg{Kubernetes} could remedy this issue, as it would make it possible to run such containers and workflows directly without alterations.

While container orchestrators are more common in cloud environments, they can and have been used on \gls{HPC} systems for container deployment, at times in conjunction with classic \gls{HPC} workload managers.
Here we review the possible integration scenarios as found in the literature \cite{wickbergSlurmVsKubernetes2022,maliaroudakisInteractiveCloudNativeWorkflows2022,lublinskyKubernetesBridgeOperator2022,lopez-huguetSeamlesslyManagingHPC2020,beltreEnablingHPCWorkloads2019}, and finally propose a new integration method.

\subsection{On-Demand Reallocation of Compute Nodes}

This setup consists of a minimal dedicated Kubernetes cluster on separate hardware, or in a suitable virtualized environment with access to the appropriate network segments. As users request nodes to run workloads on Kubernetes, the \gls{WLM} is automatically instructed to take a corresponding number of nodes offline, which are then reconfigured to run a \pkg{Kubernetes} agent (Kubelet) and connect to the \pkg{Kubernetes} cluster. \pkg{Kubernetes} can then deploy pods on these ephemeral nodes. Idling nodes must be returned automatically to the \gls{WLM} for management. Being an orchestrator for large scale infrastructure, \pkg{Kubernetes} should be well suited to this kind of dynamic reconfiguration.

\subsection{WLM in Kubernetes}

Automating the different services of a \gls{WLM} such as \pkg{Slurm} on a \pkg{Kubernetes} cluster, or any cloud orchestration service, is simple. If the containers receive privileged access to the underlying high-speed network and accelerators, the \gls{WLM} can be used to schedule \gls{HPC} jobs as in a classical \gls{HPC} system setup.

This approach does not enable running containerized workloads within the \gls{WLM}. The \gls{WLM} merely acts as a classic job scheduler, avoiding the need to rewrite workflows to directly use e.g. \texttt{mpirun} and the \gls{WLM} facilities to provision resources.
While this approach can be useful for compute centers to implement flexible multi-tenancy architectures, the \gls{WLM} needs privileged, and possibly exclusive access to hardware. Great care must thus be applied when scheduling the Pods of different tenants on the same compute nodes, or when allowing customers to obtain elevated privileges within the Pods or the \pkg{Kubernetes} instance itself.
Furthermore, any possible performance penalties incurred by the additional layer introduced must be verified.

\subsection{Kubernetes in WLM}

Initial \pkg{Kubernetes} cluster in \gls{WLM} allocation setups were evaluated on VEGA with the \pkg{CloudHypervisor} serving as a \gls{VMM}. With the advent of rootless \pkg{Kubernetes} setups, it is possible to run \pkg{Kubernetes} as a non-root user with the same technology used for running containers. This can be exploited to not only run Kubelets, but also minimal Kubernetes \pkg{K3s} and Minikube clusters, within a \gls{WLM} allocation.

In this \gls{WLM} allocation, the first node runs a minimal \pkg{Kubernetes} instance, with the other nodes running the Kubelets connecting back to this first node. The network is fully managed by the \gls{WLM}, and should not require additional configuration. While this approach permits perfect isolation between \pkg{Kubernetes} clusters started by different users, it can introduce considerable startup overhead. Until the \pkg{Kubernetes} cluster is ready, scheduling Pods or running workflows is not possible. Additional difficulties include the integration of this setup into existing user workflows, as well as the question of how users specify workloads to run. 

\subsection{Bridged Kubernetes and WLM}
\label{sec:bridging-k8s-wlm}

We identified two modalities on how to ``bridge'' \pkg{Kubernetes} and \gls{HPC} \glspl{WLM}.
The first one is via \pkg{Kubernetes} \emph{Operators}\cite{lublinskyKubernetesBridgeOperator2022}, allowing \pkg{Kubernetes} to schedule external resources.
By providing such a bridge operator, users of \pkg{Kubernetes} can use the same resource description model as for the rest of the their workloads to explicitly schedule processes on a \gls{WLM} like LFS, \pkg{Slurm} or Torque.
While this can be integrated with common workflow managers like \pkg{Kubeflow}, the drawback of this approach is the required explicit formulation in the resource description.

A more elegant approach, named KNoC and recently published by \textcite{maliaroudakisInteractiveCloudNativeWorkflows2022}, is the implementation of a virtual \pkg{Kubernetes} agent or Kubelet.
In contrast to the first, external resource management modality outlined above, a separate service acts as a regular Kubelet. It schedules Pods as jobs by starting containers using e.g. \pkg{Apptainer} within \gls{WLM} allocations, then tracks their execution and reports back.
This execution happens in an almost transparent way to the user of the \pkg{Kubernetes} cluster and to the operators of the \gls{HPC} cluster, who only have to provide the capability to run containers within their \gls{WLM}.

\subsection{Kubernetes Agent in WLM Allocation}
\label{sec:kubelet-in-wlm}

Similar to the previous scenario, this relies on a dedicated Kubernetes cluster. Rather than reallocate complete compute nodes, Kubernetes agents (Kubelets) are started as part of a WLM allocation (e.g. one Kubelet on each node). These agents then have to be able to connect back to the Kubernetes cluster to receive instructions on which containers to run. This scenario relies on the rootless-approach to run Kubernetes, and requires a compatible configuration between the \gls{WLM} and \pkg{Kubernetes}. This includes enabling version 2 of the Linux cgroups framework, cgroup delegations, and setting a suitable network configuration.

As with the KNoC approach described above, the advantage of this approach is that all the accounting information is available within the \gls{WLM}. The advantage over the KNoC, though, is the use of a fully mainline \pkg{K3s}, and therefore a standard environment for Pods to run.

This solution caters to the needs formulated at the beginning of this section~\ref{sec:kubernetes-integration}:

\begin{itemize}
    \item We want a continuously run Kubernetes cluster to schedule workflows without requiring the user to start a full Kubernetes first via the \gls{WLM}.
    \item Workloads or Kubernetes Pods should be scheduled on compute nodes \emph{within} a Slurm allocation, so as to use Slurms accounting and compute resources.
    \item Pods should run transparently on compute nodes (no changes to existing workflows).
\end{itemize}

\begin{figure}
    \includegraphics[width=\linewidth,trim=0 80pt 180pt 0,clip]{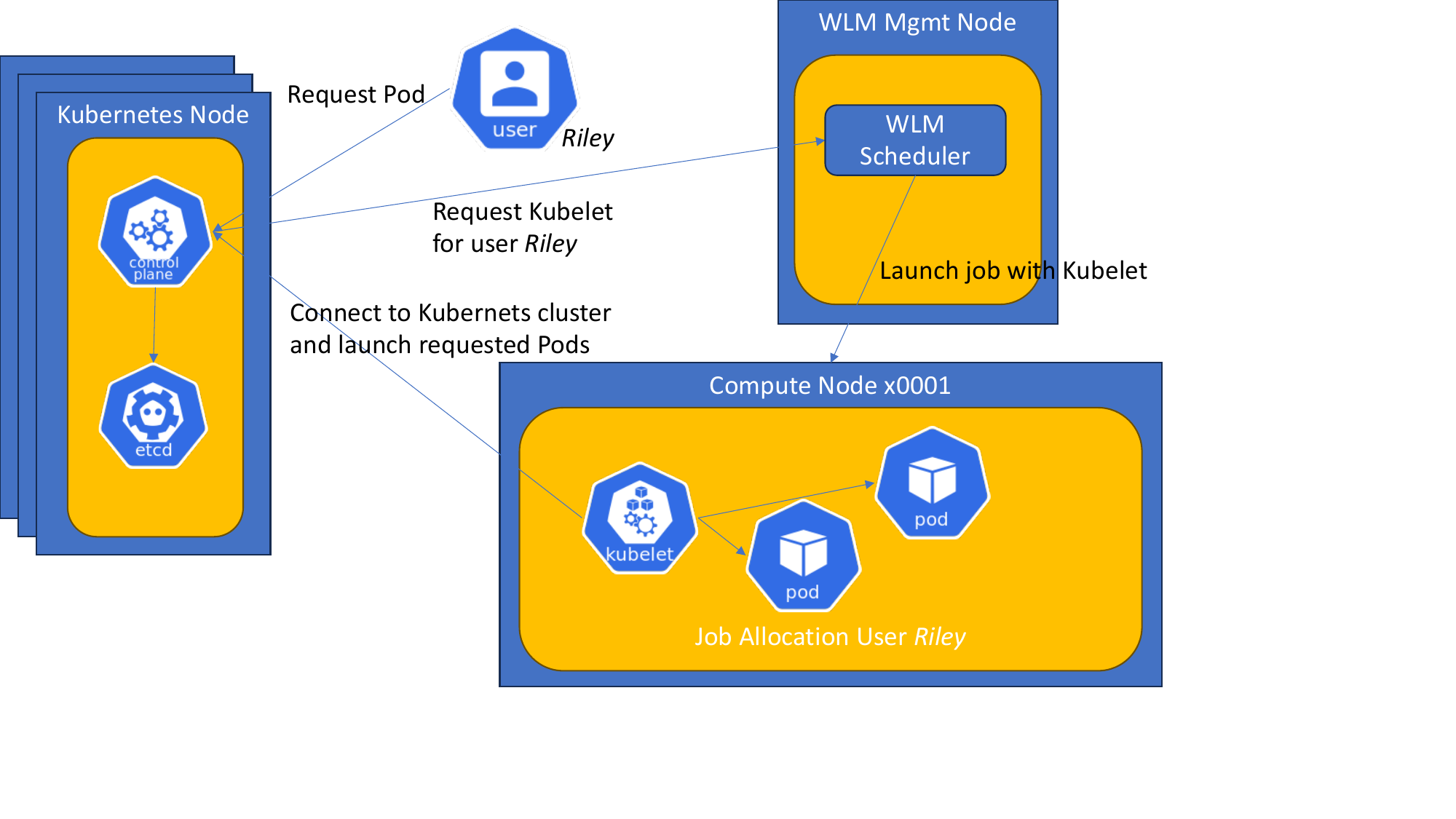}
    \caption{Principle of running Kubernetes Kubelets dynamically within a \gls{WLM} job allocation.}
    \label{fig:k3s-in-slurm}
\end{figure}

A proof-of-concept of this approach has been implemented to show the feasibility of building a \pkg{Kubernetes} cluster across the high-speed network of a compute cluster using Slingshot, and is shown in figure~\ref{fig:k3s-in-slurm}.

\subsection{Summary}

Static partitioning leads to reduced utilisation and/or a load imbalance, while dynamic partitioning, including dynamically un-/draining nodes from Slurm or Kubernetes, is cumbersome, slow and introduces disturbances to the system which may be difficult to monitor. Accounting of CPU time has to be monitored and consolidated separately.
Running the \gls{WLM} in \pkg{Kubernetes} does not provide accounting for \pkg{Kubernetes} jobs via the \gls{WLM}, and potentially introduces performance bottlenecks. Forwarding hardware resources, in particular vendor interfaces, in a secure way is non-trivial in this setting.
Conversely, running all of \pkg{Kubernetes} within a \gls{WLM} allocation leads to long startup times and requires allocations to be able to submit jobs to Kubernetes. Finally, using a Kubernetes-\gls{WLM} Operator requires a change in workflow scripts.

The only solutions satisfying the requirements are therefore the ones mentioned in section~\ref{sec:kubelet-in-wlm} and the second part of \ref{sec:bridging-k8s-wlm}. Yet for both approaches, secure multi-tenancy and transparent scheduling of multi-node Pods remain challenging, with the latter likely requiring support of the workflow manager used for scheduling.

\section{Conclusion and Outlook}
\label{sec:conclusion}

We have highlighted the specific needs of high-performance compute systems and sites, and provided an in-depth analysis of underlying operating system mechanisms. We categorized the most prominent cloud and, especially, \gls{HPC} container solutions according to the evaluated system mechanisms, providing a decision document for supercomputer operation centers. A special focus was given to security-relevant decision criteria.

From the given analysis, we infer that the \pkg{Apptainer} and Singularity family of container solutions has a unique feature set, permitting their application to a wide range of deployment scenarios, if one is willing to compromise on security. Beyond this, the HPC-specific solutions closely match what cloud-focused solutions like \pkg{Podman} provide, except for flattened filesystem images, namely unpriviled \gls{UserNS} without setuid, and using FUSE-based drivers.
With registries like \pkg{Quay} or \pkg{Dragonfly} providing \texttt{eStargz} or \texttt{EroFS} images, which can be either generated on-the-fly or uploaded in addition to the \gls{OCI} compatible layers, we assume it won't be long until these formats will be evaluated and possibly adopted for \gls{HPC} usage as an alternative to \gls{SIF}. And while Singularity has pioneered the support of encryption and signing, registry-supported solutions for both are being introduced in the cloud compute ecosystem via the \pkg{Notary}, \pkg{sigstore} and \pkg{ocicrypt} projects.
Given that cloud computing is the driving force behind container development, we expect \gls{HPC} to follow suite.

In section \ref{sec:kubernetes-integration}, we discussed why integration with Kubernetes may become relevant for supercomputing centers, and reviewed several solutions, finally proposing a new one. Two approaches were identified to address HPC needs, with their differences lying in the granularity of the scheduling, and their capability to provide a standard Kubernetes execution environment. For both approaches, multi-tenancy and security aspects must still be worked out. We did not, however, discuss the practicalities of data locality and movement. 
For container solutions, we highlighted the aspect of UID/GID mapping back to the original filesystem, indicating a manually managed and likely non-federated storage such as a cluster filesystem. The biggest advantage of API-based job scheduling via Kubernetes would be its neutrality towards the execution engine and placement, making it possible to have workflow engines place the execution of processes based on different criteria. This leaves the remaining challenge of data movement in \gls{HPC}, as cloud environments often focus on object storage, while \gls{HPC} uses cluster filesystems.

Our analysis provides the required data points to solve both containerization and abstract scheduling of applications aspects on HPC systems. What remains beyond this in regards to adaptive containerization within \gls{HPC} is the challenge of optimizing containers, selecting the most fitting optimized container and generate optimal runtime parameters for the respective target hardware in an automated fashion.

\section*{Acknowledgment}

The authors would like to thank Alfio Lazzaro, HPE for the very valuable discussions and early feedback on the draft. 

\printbibliography

\end{document}